# Imaging ferroelectric domains in multiferroics using a low-energy electron microscope in the mirror operation mode

**Salia Cherifi**[*,1,2], **Riccardo Hertel**[3], **Stéphane Fusil**[4,5], **Hélène Béa**[6], **Karim Bouzehouane**[4], **Julie Allibe**[4], **Manuel Bibes**[4] **and Agnès Barthélémy**[4]

[1] IPCMS, CNRS and UDS, 23 rue du Loess, BP43, F-67034 Strasbourg, France
[2] Institut Néel, CNRS and UJF, 25 rue des Martyrs, BP166, F-38042 Grenoble, France
[3] Forschungszentrum Jülich GmbH, IFF-9, Leo-Brandt-Str., D-52425 Jülich, Germany
[4] Unité Mixte de Physique CNRS/Thales, 1 Av. A. Fresnel, F-91767 Palaiseau and Université Paris-Sud, F-91405 Orsay, France
[5] Université d'Evry-Val d'Essonne, Bd. F. Mitterrand, F-91025 Evry, France
[6] DPMC, University of Geneva, 24 quai Ernest-Ansermet CH-1211 Geneva, Switzerland

[*] Corresponding author: e-mail cherifi@ipcms.u-strasbg.fr, Phone +33 3 8810 7218, Fax +33 3 8810 7248

We report on low-energy electron microscopy imaging of ferroelectric domains with submicron resolution. Periodic strips of 'up' and 'down'-polarized ferroelectric domains in bismuth ferrite –a room temperature multiferroic– serve as a model system to compare low-energy electron microscopy with the established piezoresponse force microscopy. The results confirm the possibility of full-field imaging of ferroelectric domains with short acquisition times by exploiting the sensitivity of ultraslow electrons to small variations of the electric potential near surfaces in the "mirror" operation mode.

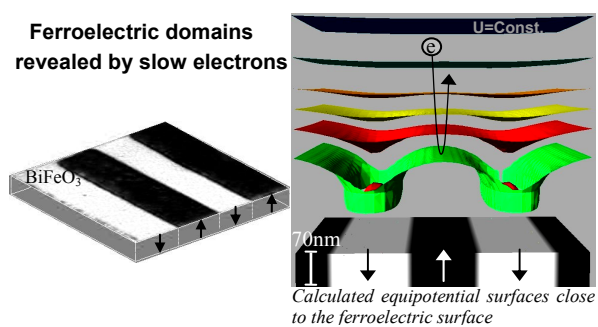

*Calculated equipotential surfaces close to the ferroelectric surface*

The manipulation of magnetic structures by electric fields and currents, rather than by magnetic fields, has recently evolved into one of the most intensively studied topics in magnetism. Since electric currents and fields are easier to generate on the nanoscale than focussed magnetic fields of well-defined strength, the fundamentally interesting electric control of magnetism is also appealing for applications like non-volatile memory devices. For such purposes, magnetoelectric multiferroics with coupled ferroelectric and ferro- or antiferromagnetic order are particularly promising materials [1]. Investigations on static and dynamic processes in these complex systems require sophisticated multi-method instruments capable of providing detailed information on the coupled magnetic and electric domain structures with short acquisition times.

Ferroelectric domain structures are routinely investigated with piezoresponse force microscopy (PFM) [2], a scanning probe technique which utilizes the converse piezoelectric effect in ferroelectric materials. PFM is usually sufficient to obtain complete information on the static ferroelectric properties of the system. The analysis of multiferroic materials, however, additionally requires the visualization of (anti-)ferromagnetic domains. Therefore complementary analysis tools are required to obtain also the magnetic properties of the magnetoelectric system. X-ray magnetic linear and circular dichroism using a photoemission electron microscope (XM(L,C)D-PEEM) was recently employed for this purpose to image magnetic domains in bismuth ferrite (BFO)-based heterostructures as a complementary method to the ex-situ PFM-study [3,4].

Here we show the possibility of imaging ferroelectric domains in BFO films with sub-micron lateral resolution by using the mirror electron microscopy mode (MEM) [5] of a non-scanning Low-Energy Electron Microscope (LEEM) [6,7]. Although MEM is an old and ripened technique -since basic mirror electron micrographs have been constructed long before LEEM- only little has been reported on the application of slow electrons for ferroelectric surface potential detection in this imaging mode since the seventies [8-11]. The use of MEM mode in nowadays advanced LEEM microscopes allows both high-lateral resolution and permits the combination of MEM with other complementary imaging modes of the LEEM system. Changing the operation mode of the microscope from the electron reflection mode to the established photoemission PEEM mode makes it possible to image also the magnetic domain structure with the same experimental setup [12], i.e., without any modification of the measurement configuration. Owing to these unique features, we anticipate that the combination of LEEM and PEEM will become the method of choice for future experimental investigations on multiferroics, in particular for studies on dynamic processes in multiferroics, such as domain wall propagation or magnetoelectric switching processes.

A 70 nm-thick antiferromagnetic-ferroelectric $BiFeO_3$ (BFO) film was grown with pulsed laser deposition on a (001)-oriented $SrTiO_3$ substrate coated with a conductive buffer layer of $La_{2/3}Sr_{1/3}MnO_3$ (LSMO). A detailed description of the growth conditions and the film characteristics can be found in Ref. [13]. 'Up' and 'down'-polarized ferroelectric strip domains were written into the BFO layer by alternate applications of negative and positive voltage (± 8 V) between the conductive PFM tip and the LSMO base electrode. Such well-defined geometric domain patterns can be easily identified in the LEEM system.

The PFM-written ferroelectric domains are imaged in the electron microscope by placing the specimen surface at a potential slightly more negative than the electron source. In this case, the incident ultraslow electrons (energy below 3 eV) do not penetrate the specimen surface as they are reflected shortly before reaching it. An electric field ***E*** applied perpendicular to the specimen surface reaccelerates the reflected electrons towards the microscope imaging column. Atomic force microscopy (AFM) was performed to ensure that the PFM-writing did not alter the surface topography of the BFO films [Fig. 1(a)]. Perfect agreement between the PFM ferroelectric domain patterns [Fig. 1(b)] and the observed MEM contrast [Fig. 1(c)] is found down to the sub-micron scale. The corresponding AFM images demonstrate that the contrast does not originate from topography. This agreement between PFM and MEM contrast is not specific to BFO. We have confirmed this result also in other systems such as Pb(Zr,Ti) layers and $BaTiO_3$ (1-2 nm) ultra-thin films (not shown). The ferroelectric contrast is easily observed at electron energies below 1 eV but the best contrast is obtained at 1-2 eV, depending on the considered surface.

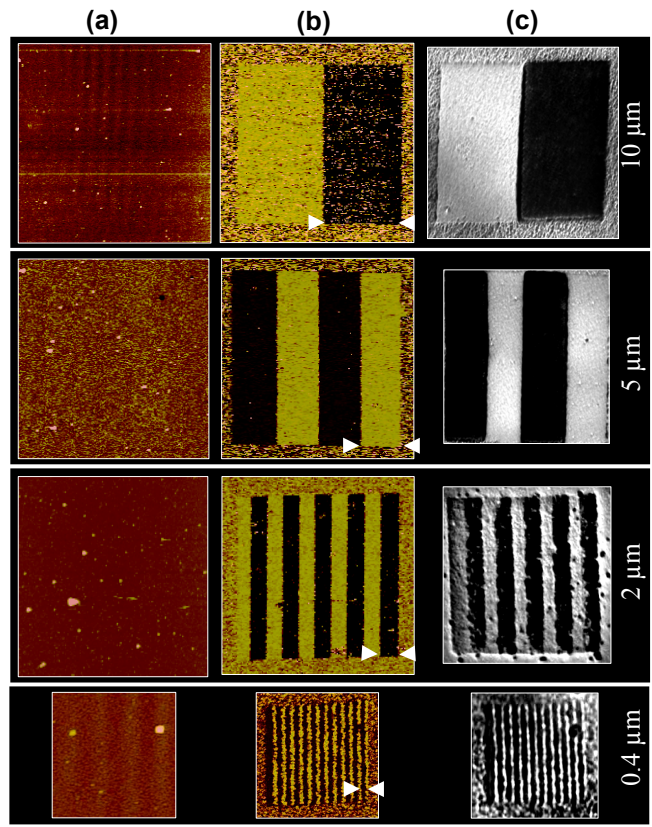

**Figure 1** Comparison between (a) AFM topography images, (b) ferroelectric domains imaged with PFM, and (c) the corresponding MEM images of 'up' and 'down'-polarized stripe domains in BFO ranging from 20 μm to 400 nm.

The MEM image formation mechanism has been described, e.g., Nepijko et al. [14] with calculations on the shift of the reflected electron trajectories induced by local microfields. These authors have also described the specimen surface potential distribution by using the current density distribution on the microscope screen [15]. Another convenient way to elucidate the ferroelectric contrast mechanism in MEM consists in representing three-dimensional equipotential surfaces (isosurfaces) $U =$ const. In the case of an electrostatically neutral and ideally flat sample exposed to an electric field oriented perpendicular to the surface, the isosurfaces are parallel to the specimen surface; each one representing an energy level. Electrons accelerated towards the sample with a given energy reach the isosurface of the corresponding energy value and are subsequently re-accelerated by the electric field, which is by definition perpendicular to the potential isosurface. Ferroelectric domains with negative and positive surface charge distributions modify the electrostatic isosurfaces near the surface leading to eggbox-like convex and concave curvatures. The resulting curved equipotential surfaces locally act as focussing or defocussing mirrors of the reflected electrons, thereby leading to dark or bright contrasts in the MEM image. The curvature of the isosurfaces is most pronounced close to the specimen surface, and flattens gradually with increasing distance. Therefore the contrast sensitivity of the mirror mode depends strongly on the distance at which the electrons are reflected. Optimum contrast is obtained with electron energies at which the reflection occurs in a region where the isosufaces display pronounced curvatures. Ideally, the point of reflection should only be about few tens of nanometres above the sample surface. An example of electrostatic potential isosurfaces calculated with the finite element method is displayed in the abstract figure.

A weaker contrast can also be obtained at higher electron energies in the standard LEEM mode (> 3 eV), where the electrons impinge the sample and interact directly with the material. In this case the contrast arises from the electric potential variations experienced by the electrons inside the ferroelectric material and from field distortions close to the domain boundaries which modify the trajectories of the backscattered electrons. In addition, when the energy of the incident electron beam is tuned to the low-energy crossover of the secondary electron yield of the material, the emitted secondary electrons can be used to image the surface. Low-resolution ferroelectric contrast can also be obtained in this secondary electron emission mode due to the polarization dependence of the emission threshold, as previously demonstrated using UV-PEEM [16]. Such contrast can also be obtained in XMLD-PEEM mode, since x-ray linear dichroism is not only sensitive to the antiferromagnetic order but also to the electric polarization [4]. In contrast to this, the mirror electron mode is exclusively sensitive to the ferroelectric contribution [Fig.2 (a)]. In the secondary electron emission mode, the lateral resolution is however about two to three times lower than in the MEM reflection mode [Fig.2 (a,b)] because of the large solid angle of electron emission and the energy spread of the emitted electrons, leading to spherical and chromatic aberrations. From line scans across opposite ferroelectric domains we could deduce that the lateral resolution in the MEM mode is better than 15 nm. While ultimate resolution in MEM has not yet been quantitatively demonstrated, a lateral resolution of few nanometres has been predicted by Remfer and Griffith [17].

Imaging ferroelectric domains with slow electrons offers several advantages: (a) MEM is a 'zero-impact' technique where charging effects are minimized since the electron-probe does not penetrate or even reach the specimen surface; (b) No particular sample preparation is needed for imaging (such as thinning procedures in transmission microscopy) and the base electrode (necessary for PFM) is not required for MEM, thereby allowing also the analysis of bulk materials; (c) Short acquisition times in the quasi-static mode (e.g., during *in situ* growth) and fast time-resolved measurements in the pump-probe scheme are both possible [18]; (d) In addition to ferroelectric domains, spectroscopic-LEEM instruments [12] make it also possible to image magnetic domains with the same setup in static (Fig.2b) or time-resolved X-PEEM mode. High-resolution imaging of ferromagnetic domains is also accessible in the LEEM system when using a spin-polarized electron source in the SP-LEEM mode.

In conclusion, we have demonstrated the imaging of ferroelectric domains with low-energy electron microscopy in the mirror mode down to the sub-micron scale. This setup is particularly versatile for the study of magnetic multiferroics and should allow for fast imaging of both ferroelectric and magnetic domains. The possibility of applying in situ magnetic and electric fields further enhances the potential of this instrumentation for future investigations on the magnetoelectric coupling and to dynamic processes in multiferroics.

**Acknowledgements** The authors thank the team of Nanospectroscopy beamline for the assistance during the experiment at Elettra. SC tanks A. Fraile-Rodriguez and F. Nolting for the support during the first exploratory experiment at the SLS.

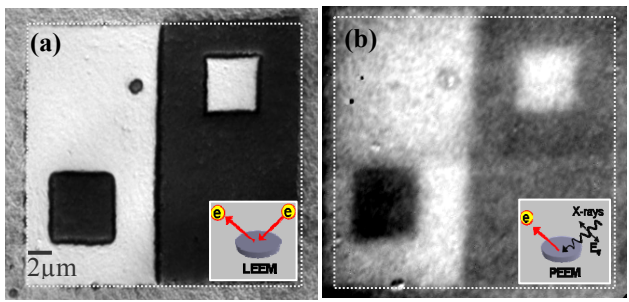

**Figure 2** (a) Low-energy electron microscopy (MEM-mode) image of PFM-written perpendicular ferroelectric domains and (b) the corresponding XMLD-PEEM image displaying the local electron emission yield obtained with X-rays tuned at the FeL$_3$ multiplet in BFO. The linear polarization vector E is oriented parallel to the film plane. Spectroscopic LEEM can be used to image (a) reflected electrons or (b) x-ray induced photo-emitted electrons as illustrated in the sketches.